# REALIZATION AND DESIGN OF A PILOT ASSIST DECISION-MAKING SYSTEM BASED ON SPEECH RECOGNITION

Jian ZHAO, Hengzhu LIU, Xucan CHEN and Zhengfa LIANG

School of Computer, National University of Defense Technology, 410073 Changsha, China
zhaojian9014@gmail.com

## ABSTRACT

*A system based on speech recognition is proposed for pilot assist decision-making. It is based on a HIL aircraft simulation platform and uses the microcontroller SPCE061A as the central processor to achieve better reliability and higher cost-effect performance. Technologies of LPCC (linear predictive cepstral coding) and DTW (Dynamic Time Warping) are applied for isolated-word speech recognition to gain a smaller amount of calculation and a better real-time performance. Besides, we adopt the PWM (Pulse Width Modulation) regulation technology to effectively regulate each control surface by speech, and thus to assist the pilot to make decisions. By trial and error, it is proved that we have a satisfactory accuracy rate of speech recognition and control effect. More importantly, our paper provides a creative idea for intelligent human-computer interaction and applications of speech recognition in the field of aviation control. Our system is also very easy to be extended and applied.*

## 1. INTRODUCTION

Speech recognition is a technology which is used to implement an appropriate control through correctly identifying and judging the speech characteristics and connotation [1]. In recent years, the applications of speech recognition technology in the fields like human-computer interaction have become more and more popular and challenging. As a very important technological progress, the pilot assist decision-making based on speech recognition can reduce burden on pilot, lower operating risk, and improve cockpit human-machine interface [2]. However, domestic application of speech recognition is still in a big blank at present. It's a great help to carry out pre-research in time, to understand and to master the technology, to overcome the application difficulties for improving the application level of our aviation control technologies.

Currently, DSP (Digital Signal Processor) chips are mainly applied to speech recognition. But they are generally more expensive, more complex and harder to be extended and applied [3]. The system proposed in our paper is realized with the HIL aircraft simulation platform and the 16-bit microcontroller SPCE061A. SPCE061A acts as the central processor for digital speech recognition to achieve better reliability and higher cost-effect performance. Technologies of LPCC and DTW are applied for isolated-word speech recognition to gain a smaller amount of calculation and a better real-time performance. Besides, we adopt the PWM regulation technology to effectively regulate each control surface by speech, and thus to assist the pilot to make decisions.

The rest of the paper is organized as follows: algorithm of speech recognition is described in detail in the second part; hardware structure and software design of a pilot assist decision-making system based on speech recognition are respectively elaborated in Part III and Part IV; the performance of our whole system is evaluated in Part V; in the last part, we draw a summary and look forward to the future work.

## 2. ALGORITHM OF SPEECH RECOGNITION

As can be seen in Figure 1, speech recognition is essentially a kind of pattern recognition, which consists of basic units such as pretreatment, A/D conversion, endpoint detection, feature extraction and recognition judgment, [4] etc. According to the basic principle of pattern recognition, by comparing the pattern of the unknown speech with the reference pattern of the known, we can obtain the best matched reference pattern, namely, the result of recognition.

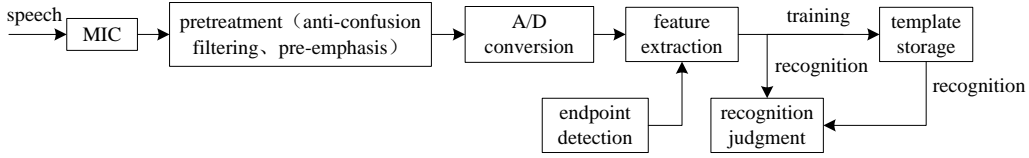

Figure 1. The basic structure of the speech recognition system

### 2.1. Endpoint Detection

Endpoint detection means using the digital processing technology to identify the start point and the end point among all kinds of paragraph (phonemes, morphemes, words, syllables, etc.) in the speech signal. The most commonly used method in endpoint detection is based on short-term energy and short-term zero-crossing rate [5, 6].

Short-term energy is defined as follows:

$$E_n = \sum_{m=-\infty}^{\infty} \left[ x(m) \cdot w(n-m) \right]^2 \quad (1)$$

Here $E_n$ reflects the law of the amplitude or energy of the voiced/unvoiced frames in the speech signal changing slowly over time [7]. According to the change of $E_n$, we can roughly judge the moment when voiced frames turn into unvoiced ones and the unvoiced frames turn into voiced ones. $E_n$ is very sensitive to the high-level signal because the square of it was used when calculated formula (1). So in practice, we also use the following two types of definition:

$$E_n = \sum_{m=-\infty}^{\infty} \left| x(m) \cdot w(n-m) \right| \quad (2)$$

$$E_n = \sum_{m=-\infty}^{\infty} \log^2 \left[ x(m) \cdot w(n-m) \right] \quad (3)$$

Short-term zero-crossing rate is defined as follows:

$$Z_n = \sum_{m=-\infty}^{\infty} \left| \operatorname{sgn}\left[ x(n) \right] - \operatorname{sgn}\left[ x(n-1) \right] \right| \cdot w(n-m) \quad (4)$$

sgn[•] is the sign function:

$$\operatorname{sgn}(n) = \begin{cases} 1, & x(n) \geq 0 \\ 0, & x(n) < 0 \end{cases} \quad (5)$$

$$w(n) = \begin{cases} \dfrac{1}{2N}, & 0 \leq n \leq N-1 \\ 0, & else \end{cases} \quad (6)$$

$Z_n$ means the total number that the speech signal changes from positive to negative and from negative to positive per unit time [8]. According to $Z_n$, we can roughly obtain the spectral characteristics of the speech signal to distinguish the voiced/unvoiced frames and whether there's speech or not.

A two-stage judgment method is usually adopted in endpoint detection based on $E_n$ - $Z_n$. As can be seen in Figure 2, firstly, select a relatively high threshold M1 according to the outline of $E_n$, in most cases, M1 is below $E_n$. In this way, we can do a rough judgment: the start point and the end point of the speech segment are located outside of the interval corresponding to the intersection points of envelops of M1 and $E_n$ (namely, outside of segment AB). Then determine a relatively low threshold M2 based on the average energy of the background noise, and search from point A to the left, point B to the right, find out the first two intersection points C and D of the envelop of $E_n$ and the threshold M2, so segment CD is the speech segment determined by the dual-threshold method according to $E_n$. From above we just finished the first stage of judgment, then turn to the second: use $Z_n$ as the standard, and search from point C to the left, point D to the right, then find out the first two points E and F which are lower than threshold M3, so they are the start point and the end point of the speech segment.

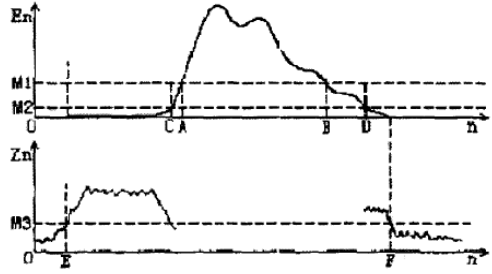

Figure 2. Endpoint detection based on $E_n$-$Z_n$

## 2.2. Feature Extraction

Feature extraction is a crucial step in speech recognition. If the speech features were effectively extracted, it is easier for us to distinguish among different categories in the feature space. Compared to the other speech features, linear predictive cepstral coding (LPCC) can effectively represent the speech feature, including channel characteristics and auditory features. LPCC has an excellent distinguishability, speed and accuracy. Besides, LPCC can effectively guarantee the real-time performance of speech recognition. LPCC is calculated based on linear prediction coefficient (LPC) characteristics:

$$\begin{cases} c(1) = a(1) \\ c(n) = \sum_{k=1}^{n-1} \left[1 - \frac{k}{n}\right] \cdot c(n-k) \cdot a(k) + a(n) \end{cases} \quad (7)$$

$c(n)$ is the coefficient of LPCC, (n=1,2,…,p); p is the feature model order, most channel models of the speech signal can be sufficiently approximated when we take p=12; $a(k)$ is the linear prediction coefficient (LPC) characteristics.

## 2.3. Recognition Judgment

We apply DTW algorithm, which is the commonly used identification method in speech recognition, to the recognition judgment part.

The basic idea of DTW is to find out the phonological characteristics of the speech signal and compare the distance，that is, to find out the differences (characteristic differences) between the frame characteristics in chronological order; And then accumulate characteristic differences included in phonological features and divided by the whole characteristic difference of the entire pronunciation. At last, we get the relative cumulative characteristic difference. Thus, despite the different articulation rate, the relative cumulative characteristic differences of the phonological characteristics are basically the same. Specific algorithm is as follows:

(I) normalize the feature data $L(i, j)$ (the coefficient of LPCC) per frame, and get $S(i, j)$, the characteristic difference between two adjacent frames is:

$$t(j) = \sum_{i=1}^{c} |s(i, j) - s(i, j+1)| \qquad (8)$$

The average characteristic difference is:

$$t = \frac{1}{N-1} \sum_{j=1}^{N-1} t(j) \qquad (9)$$

$N$ is the number of speech frames.

(II) Check $t(j)$ from back to front, remove the ones that larger than the average characteristic difference, until it's less than the average characteristic difference, so as to remove the end part that contains less semanteme. The number of data frames reduced to $N'$. Assume the cumulative characteristic difference threshold is:

$$\Delta = \frac{1}{M} \sum_{j=1}^{N'} t(i) \qquad M \leq N'-1 \qquad (10)$$

$M$ is the number of reserved key frames. Usually we take $M=8$ for isolated character sound, and $M=16$ for double isolated words.

(III) Pick out the key frames: the first frame must be chosen, then plus $t(i)$ in turn, the frame greater than $\triangle$ is another key frame, until $M$ key frames are picked out.

(IV) Piecewise linearization, and take the average of the characteristic differences between two key frames as the last speech eigenvector. In the process of training, save these feature vectors as templates. And in the process of recognition, match the speech signals and the templates, and calculate the distances, the minimum one within the scope of the distance threshold is the final recognition result.

## 3. HARDWARE STRUCTURE

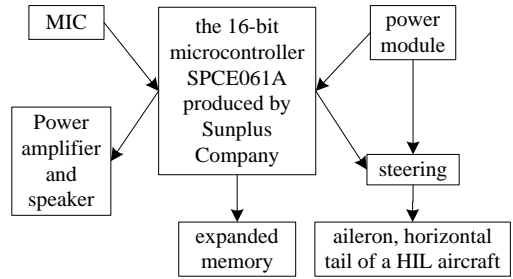

Figure 3. Hardware block diagram of the system

As can be seen in Figure 3, the hardware structure of the pilot assist decision-making system based on speech recognition mainly includes a microprocessor circuit module based on a 16-bit microcontroller SPCE061A produced by Sunplus Company, expanded memory, audio circuit module, power module, steering and executive components.

SPCE061A contains multiple A/D converters, dual-channel 10-bit D/A converters and an online simulation circuit ICE interface. Besides, SPCE061A has the advantages of smaller size, higher integration, better reliability and cost-effective performance, easier to be expanded, stronger interrupt processing ability, more efficient instruction system and less power consumption, etc. than DSP chips [9]. In order to achieve real-time processing of speech signal, the whole hardware system is divided into the following several parts:

(I) Extraction, training, and judgment of speech features: we use speech processing and DSP functions of SPCE061A to pre-emphasis on the input speech digital signals, then cache and extract feature vectors, create templates under the function of training, and make judgment under the function of recognition.

(II) Acquisition of speech signals: SPCE061A has a microphone amplifier and single-channel speech A/D converters with the function of automatic gain control so that we could save much front-end processing hardware, simplify the circuit, and improve the stability. Connect the microphone to the anti-aliasing filter and access to the channel, and then complete sampling of the 10-bit 8 kHz signal.

(III) Expand memory: we need to expand a flash memory of 32 KB as the data memory because processing of speech signals requires a large amount of data storage. The storage space is divided into 4 parts: templates storage area is used to store isolated-word feature templates, and the number of stored templates (namely, the number of identifiable vocabulary) is determined by the size of the storage area; speech signal temporary storage area is used to store 62 frames of data of each speech signal to be identified; intermediate data storage area contains a 2 KB SRAM, and it's used to store the intermediate computation, such as background noise characteristics and intermediate templates produced in the process of training, etc. the prompt speech information storage area is used to store function prompts speech and recognition response speech, etc. so as to facilitate human-computer interaction. Input of this part of speech signals can be achieved by the software wave_press provided by Sunplus Company.

(IV) Control input: consist of three keys, they are function switch key (recognition/study), function confirms and exit key, template revise and select key (select and modify a template). Through these three keys, we could realize the human-computer interaction of FS (Function Selection).

(V) Output: include a speaker output part and a control output part. The speaker is connected to a dual-channel 10-bit D/A converter with the function of audio output, and is used to output prompts speech and recognition response speech. The control output part is used to output a control signal through I/O interface, and then adjust corresponding steering, change flight attitudes, to realize assistant decision-making after having recognized speech instructions.

## 4. SOFTWARE DESIGN

Our system's software implement is developed in the integrated developing environment IDE3.0.4 of SPCE061A based on C language, which mainly includes three parts: main program, interrupt handling routine and function module subroutine. We will introduce the three parts in detail as follows.

### 4.1. Main Program

As can be seen in Figure 4, the processes of the main program are divided into initialization, training and recognition. The training and recognition of the speaker-dependent speech could be accomplished by calling related functions, and the corresponding operations could be performed according to the results of recognition.

(I) Initialization: The system collects 10 frames of background noise data at first after power on reset, extracts features of En and Zn after pre-emphasis, and then determines the threshold value as the basis to identify the start point and the end point.

(II) Training: enter by the function switch key, and prompt "now is the training function". And then prompt "now modify the first template", select the template to be modified by the template revise and select key, after per click it turns to the next template. And then prompt "speech input at the first time", you are asked to input 4 times here in order to ensure the accuracy of the template. Extract the feature vectors and temporary store the template. Only after 4 times all succeeded would it prompt "successfully modified". Otherwise, data won't be retained, and template won't be modified, neither. If one process lasted for more than 10s, it would prompt "Quit the training function" and do not make any changes.

(III) Recognition: The microcontroller constantly sample the outside signal, and save 10 frames of speech data to judge the start point; and then sample 52 frames of speech data to determine the end point. Handle the error if there's no end point. After that, calculate LPCC of each frame, and use LPCC and DTW to get the eigenvectors of isolated words in that speech segment. Compare them with the templates, if the distance is within the specified threshold, select the template with the minimum distance as the result. At the meantime, a corresponding response is made. However, if the distance is beyond that threshold, handle the error.

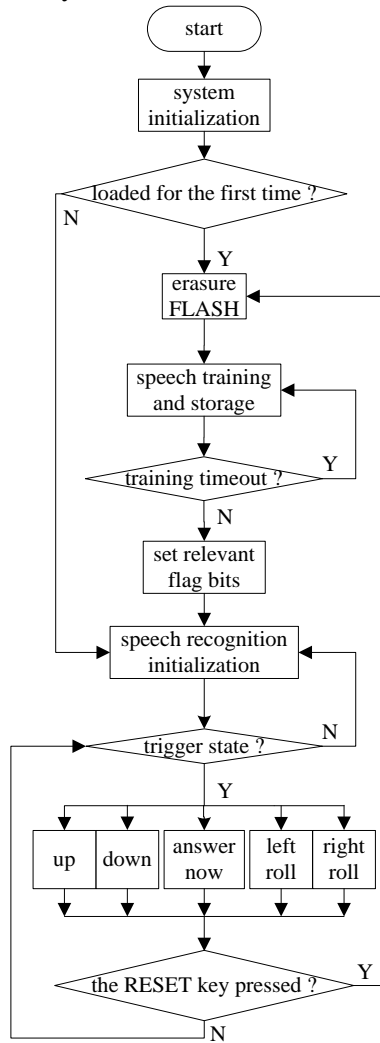

Figure 4. Software flow of the main program

## 4.2. Interrupt Handling Routine

As can be seen in Figure 5, A/D conversion results are read by the interrupt handling routine periodically and deposited in the buffer. The speech signal of MIC channel is the input of A/D. Interrupts are generated by speech recognition and playback TMA_FIQ interrupt sources, and judged by the flag bit whether it's speech playback or speech recognition [10, 11]. Functions written in the process of speech recognition are: start point judgment function, end point judgment function, LPC function, LPCC function, characteristic differences piecewise linear dynamic distribution function, judgment function, error function, and the upper functions, feature extraction functions constructed by these sub-functions.

## 4.3. Function Module Subroutine

The function module subroutine includes the functions of up, down, left roll, right roll, reset and output of PWM wave, etc. In each function, corresponding operation is realized by configuring I/O output to provide related signal to the circuit of steering.

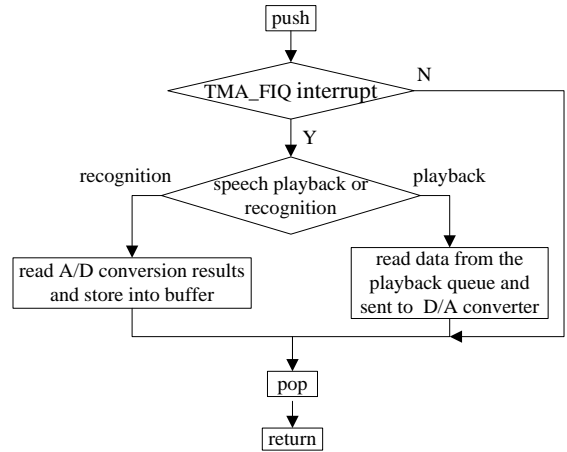

Figure 5. Software flow of the interrupt handling routine

The steering is a position servo actuator, and applies to those control systems in need of the ever-changing and maintaining angle [12]. The control of steering usually needs a time-base pulse around 20ms, and the high level part of the pulse is generally 0.5ms-2.5ms. Take the servo of 180 degrees angle for example, the corresponding control relationships are shown in Table 1.

Table 1. Control relationship between pulse and pivot angle of steering.

| Variables | Negative | Negative | Zero | Positive | Positive |
|---|---|---|---|---|---|
| High level | 0.5ms | 1.0ms | 1.5ms | 2.0ms | 2.5ms |
| Angle | 0 | 45 | 90 | 135 | 180 |

In the drive program of steering, we take the angle of 90 degrees corresponding to 1.5ms as the initial position of the system, and realize the zero declination of control surface through the reset function. By changing the duty ratio of the output PWM wave in the functions of up, down, left roll and right roll, to control the positive and negative angle of steering, and thus, to control the aircraft flight attitudes. Figure 6 shows the output of PWM wave when the high level part is 1.5ms.

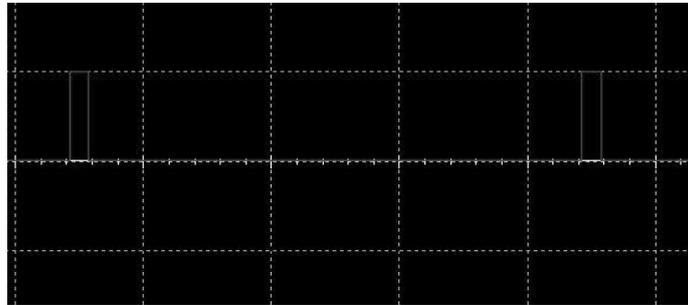

Figure 6. Output of PWM wave when the high level part is 1.5ms

## 5. SYSTEM PERFORMANCE ANALYSIS

For the same training template, let the speaker dependent and the speaker independent respectively test the system based on a HIL aircraft simulation platform, 20 times test for each command and 100 times test for each group. The results show that: the recognition rate of speaker dependent has reached more than 95.3%, the recognition rate of the speaker-independent A is 81.2%, the recognition rate of the speaker-independent B is 85.7%; then select male speech as the template, and use female speech test the system, the recognition rate is 54.5%. From the results we know that the recognition rate of speaker dependent is higher than speaker independent. Besides, a higher recognition rate could be obtained if a more precise algorithm was taken during template matching.

In the aspect of aircraft flight attitudes control, we used SOLIDWORKS design and simulate the movements of the corresponding control surface. As can be seen in Figure 7, control parts move the front end of the aircraft horizontal tail upward after system recognized the instruction "down" of the pilot.

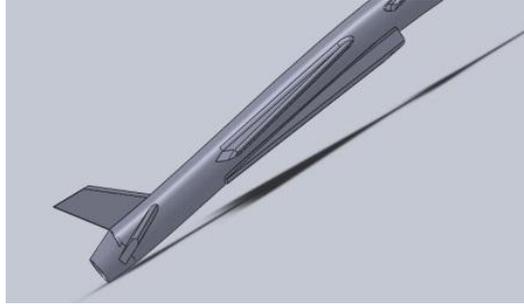

Figure 7. Aircraft down

As can be seen in Figure 8, control parts move the trailing end of the aircraft left aileron upward and the right aileron downward after system recognized the instruction "left roll" of the pilot.

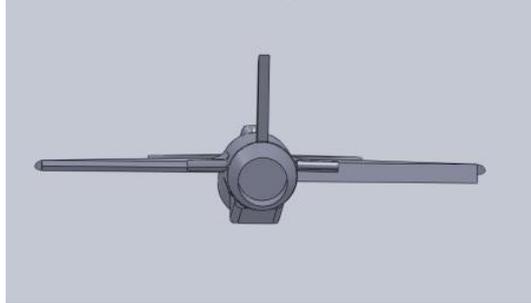

Figure 8. Aircraft left roll

By trial and error, the movements of each control surface under the corresponding speech instructions are in line with expectations and the overall performance is better than that gained by other methods.

## 6. CONCLUSIONS

As a very important technological progress, the pilot assist decision-making based on speech recognition can reduce burden on pilot, lower operating risk, and improve cockpit human-machine interface [13]. However, domestic application of speech recognition is still in a big blank at present. It's a great help to carry out pre-research in time, to understand and to master the technology, to overcome the application difficulties for improving the application level of our aviation control technologies.

The system proposed in our paper is realized with the HIL aircraft simulation platform and the 16-bit microcontroller SPCE061A. SPCE061A acts as the central processor for digital speech recognition to achieve better reliability and higher cost-effect performance. And an artificial intelligence system is introduced in the control system of aircraft to achieve more flexible control and better human-computer interaction. Besides, speech recognition is optimized by certain mechanical structures and algorithms. Speech features and recognition methods fit for speaker-dependent isolated word are selected to achieve faster processing speed and higher recognition rate, so as to meet the needs of real-time speech recognition [14, 15]. Our system made the best advantages of speech control and realized a system for assisting the pilot to make decisions. By trial and error, it is proved that we have a satisfactory accuracy rate of speech recognition and control effect.

## ACKNOWLEDGEMENTS

This work is supported by the Doctor Program Foundation of Education Ministry of China, under grant No. 20114307130003. We thank Haibin DUAN, Hongyu YAN and Chi MA from Beijing University of Areonautics and Astronautics in China for their valuable comments to improve this paper.## REFERENCES

[1] Lavner Y, Gath I, Rosenhouse J.: The effects of acoustic modificationson the identification of familiar voices speaking isolated vowels. Speech Communication (2000).

[2] Chu M K, Sohn Y S.: A user friendly interface operated by the improved DTW method. In: 10th IEEE International Confe-rence on Fuzzy Systems (2002).

[3] LEE C H.: On Automatic Speech Recognition at the Dawn of 21th Century. IEICE TRANS, INF & SYST (2003).

[4] Lee, D. Hyun, etc.: Optimizing feature extraction for speech recognition. IEEE Transactions on Speech and Audio Processing (2003).

[5] X. Huang, A. Acero, H. W. Hon.: Spoken language processing-a guide to theory, algorithm, and system development (2003).

[6] Zbancioc M, Costin M.: Using neural networks and LPCC to improve speech recognition signals. Proceedings of the International Symposium on Circuits and Systems (2003).

[7] SU JAY P, RH ISH IKESH L, SIDDH ARTH V.: On design and implementation of an embedded automatic speech recognition system. IEEE Transactions on VLSI Design (2004).

[8] Turner CW, Gantz BJ, Vidal C, et al.: Speech recognition in noise for cochlear implant listeners: benets of residual acoustic hearing. The Journal of the Acoustical Society of America (2004).

[9] Osamu Segawa, Kazuya Takeda, Fumitada Itakura.: Continuous speech recognition without end-point detection. Electrical Engineering (2006).

[10] Abu Shariah, Mohammad A. M. Ainon, Raja N. Zainuddin, Roziati. Khalifa, Othman O.: Human computer interaction using isolated-words speech recognition technology. 2007 International Conference on Intelligent and Advanced Systems, ICIAS 2007 (2007).

[11] J.H.L Hansen, L. M. Arslan.: Robust feature estimation and objective quality assessment for noisy speech recognition using the credit card corpus. IEEE Trans. Speech Audio Processing (2009).

[12] Je-Keun Oh, Giho Jang, Semin Oh, Jeong Ho Lee, Byung-Ju Yi, Young Shik Moon, Jong Seh Lee, Youngjin Choi.: Bridge inspection robot system with machine vision. Automation in Construction (2009).

[13] Rabiner. L. R.: An algorithm for determining the endpoints of isolated utterance. The Bell System Technical Journal (2010).

[14] Mayumi Beppu, Koichi Shinoda, Sadaoki Furui.: Noise Robust Speech Recognition based on Spectral Reduction Measure. APSIPA ASC (2011).

[15] Microsoft Speech SDK 5.1 Help. http://www.Microsoft.com.

**Authors**

Jian ZHAO received the B.S. degree in Automatic Control and Information Technology from Beijing University of Aeronautics and Astronautics of China in 2012, and he is currently working toward the M.S. degree in Computer Science and Technology in National University of Defense Technology. His research interests are in the area of computer system architecture, software radio communication, and signal processing with huge data.

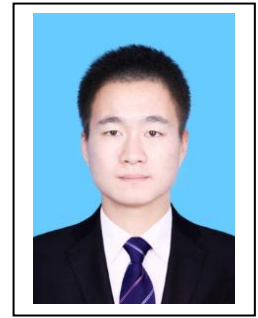

Hengzhu LIU received the Ph.D. degree in Computer Science and Technology from National University of Defense Technology of China in 1999. Currently, he is a professor and vice leader with the Institute of Microelectronics and Micro-processors of computer school in National University of Defense Technology. His research interests include processor architecture, micro-architecture, memory architecture, system on chip (SoC), VLSI signal processing and high-performance digital signal. He is a member of IEEE, IEICE and China Computer Federation.

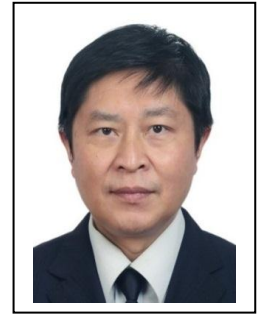

Xucan CHEN received the Ph.D. degree in Computer Science and Technology from National University of Defense Technology of China in 1999. Currently, she is a professor of computer school in National University of Defense Technology. Her research interests include computer system architecture, micro-architecture, software radio communication, VLSI signal processing and high-performance digital signal.

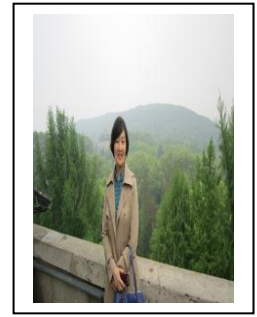

Zhengfa LIANG received the M.S. degree in Electrical Science and Technology from National University of Defense Technology of China in 2012. He is currently working toward the Ph.D degree in Electrical Science and Technology in National University of Defense Technology. His research interests are in the area of wireless communication, parallel processing architectures, and circuits.

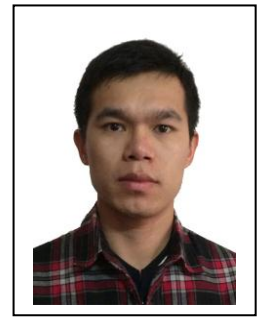